\documentclass{PoS}

\usepackage{epsfig}
\usepackage{latexsym}
\usepackage{graphicx}
\usepackage{bm}
\usepackage{longtable}

\def\lsim{\raise0.3ex\hbox{$<$\kern-0.75em\raise-1.1ex\hbox{$\sim$}}}
\def\gsim{\raise0.3ex\hbox{$>$\kern-0.75em\raise-1.1ex\hbox{$\sim$}}}
\def\simgt{\rlap{\lower 3.5 pt\hbox{$\mathchar \sim$}}\raise 1.0pt \hbox {$>$}}
\def\simlt{\rlap{\lower 3.5 pt\hbox{$\mathchar \sim$}}\raise 1.0pt \hbox {$<$}}

\title{Calculation of Helium nuclei in quenched lattice QCD}

\ShortTitle{Calculation of Helium nuclei in quenched lattice QCD}

\author{Takeshi Yamazaki
\hspace{-1mm}{\footnote{Present address: Kobayashi-Maskawa Institute for the Origin
 of Particles and the Universe, Nagoya University, Naogya, Aichi 464-8602, 
Japan}}
\ for PACS-CS Collaboration
\\ \\
Center for Computational Physics, University of Tsukuba,
Tsukuba, Ibaraki 305-8577, Japan\\
        \email{yamazaki@kmi.nagoya-u.ac.jp}
}

\abstract{
We present results for the binding energies for $^4$He and $^3$He nuclei 
calculated in quenched lattice QCD at the lattice spacing of $a =0.128$ fm
with a heavy quark mass corresponding to $m_\pi = 0.8$ GeV.
Enormous computational cost for the nucleus correlation functions is
reduced by avoiding redundancy of equivalent contractions 
stemming from permutation symmetry of protons or neutrons in the 
nucleus and various other symmetries.
To distinguish a bound state from an attractive scattering state,
we investigate the volume dependence of the energy difference 
between the ground state energy of the nucleus channel 
and the free multi-nucleon states by changing the
spatial extent of the lattice from 3.1 fm to 12.3 fm.
A finite energy difference left in the infinite spatial volume limit
leads to the conclusion that the measured ground states are bounded.
It is also encouraging that the measured binding energies and the experimental 
ones show the same order of magnitude. 
}

\FullConference{The XXVIII International Symposium on Lattice Field Theory, Lattice2010\\
		June 14-19, 2010\\
		Villasimius, Italy}

\begin{document}

\section{Introduction}

The atomic nuclei have been historically treated as collections of protons and 
neutrons.  The great success of the nuclear shell model since 
1949~\cite{{Mayer:1949xx},{Haxel:1949xx}}, 
explaining the nuclear magic numbers and detailed spectroscopy, has established 
that protons and neutrons are very good effective degrees of freedom 
at the nuclear energy scale of a few MeV.  
Nonetheless, 60 years later, we know for certain that protons and neutrons 
are made of quarks and gluons whose laws are governed by QCD.  
It is a great challenge to quantitatively understand the structure 
and property of known nuclei based on the first principle of QCD. 
This direct approach will be more important and indispensable if we are to 
extract reliable predictions for experimentally unknown nuclei in the 
neutron rich regions of the nuclear chart. 
In this article we address the fundamental question in the research in this 
direction, namely the binding energies of nuclei.

Lattice QCD study of multi-baryon states goes back a long time, starting with H 
dibaryon~\cite{Jaffe:1976yi} in the 80's~\cite{Mackenzie:1985vv} and deuteron 
in the early 90's~\cite{Fukugita:1994na}.  
More recently, exploration of three baryon states  began~\cite{Beane:2009gs}.  
So far, however, there is no established evidence supporting bound state 
formation in these channels.  An exception is a model study 
in the strong coupling limit of lattice QCD~\cite{deForcrand:2009dh}.

We attempt to go a step further in mass number and examine the helium nuclei, 
especially $^4$He with the mass number $A=4$. 
Besides the obvious physical interest as 
the first natural element beyond hydrogen, it is also the system where 
technical difficulties of fermion contractions specific to nuclei with 
a large mass number appear in a non-trivial way. On the other hand, 
the binding energy drops down to a large value of $\Delta E = 28.3$ MeV 
for the $^4$He nucleus, making us hopeful that observing the bound state nature 
might be easier than the lighter nuclei. 

The organization of this article is as follows.
In Sec.~\ref{Historical perspective} we review previous studies 
for bound states in multi-baryon systems from lattice QCD.
The computational issues with studies of multi-baryon states 
and their solutions employed in this work 
are briefly explained in Sec.~\ref{Computational issues with nuclei}.
The simulation details and the results for the $^4$He and $^3$He channels
are presented in Sec.~\ref{A quenched calculation of Helium nuclei}.
A brief summary and a look toward future are given in Sec.~\ref{Toward further progress}.
The results in this article have been reported in Ref.~\cite{Yamazaki:2009ua}.

\section{Historical perspective}
\label{Historical perspective}

Bound states in multi-baryon systems have been investigated by several 
studies in lattice QCD.
For systems with two baryons, the first study 
was the search for the H dibaryon.
According to Jaffe~\cite{Jaffe:1976yi} the H dibaryon having strangeness 
$S=-2$ and isospin $I=0$ channel was expected to 
have a large binding energy of $O(100)$ MeV.   
Most of the quenched lattice QCD studies~\cite{{Mackenzie:1985vv},{Iwasaki:1987db},
{Pochinsky:1998zi},{Wetzorke:1999rt},{Wetzorke:2002mx}} 
concluded that the H dibaryon bound state does not exist.
Recently NPLQCD Collaboration observed a small, negative energy shift,
$E_{\Lambda\Lambda} - 2 m_{\Lambda} = -4.1(1.2)(1.4)$ MeV~\cite{Beane:2009py},
in this channel.
They concluded, however, that the evidence is not strong enough to 
establish the existence of the H dibaryon, and that further study is 
necessary with different volumes.  The latter point is related to the 
computational problem of the nucleus calculation, which we will discuss in the next section.

\begin{figure}[!t]
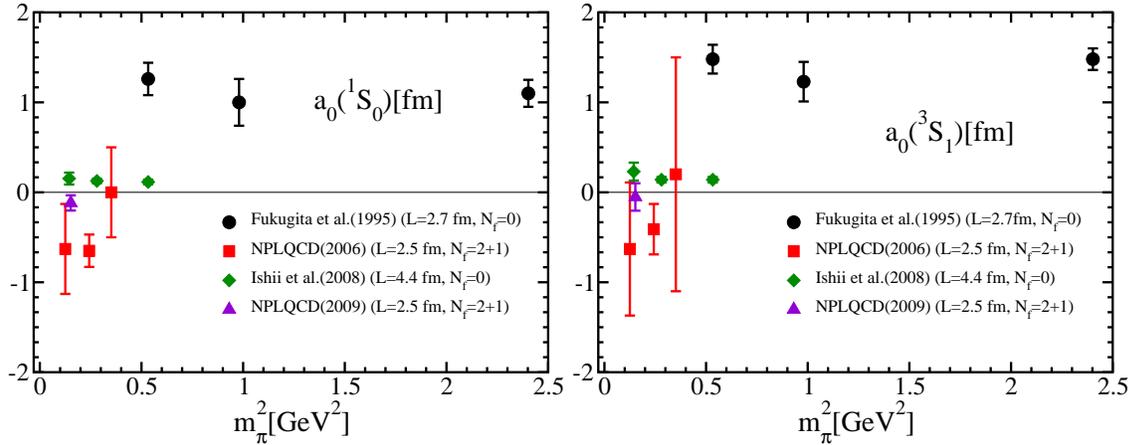

\includegraphics*[angle=0,width=0.49\textwidth]{Figs/nn_1s0_0.eps}
\includegraphics*[angle=0,width=0.49\textwidth]{Figs/nn_3s1_0.eps}
\caption{
Scattering length for $^1S_0$ (left) and $^3S_1$ (right) channels.
Circle, square, diamond, and triangle denote results for
quenched~\cite{{Fukugita:1994na},{Fukugita:1994ve}},
mixed action~\cite{Beane:2006mx},
two-nucleon wave function~\cite{Ishii:2006ec},
and anisotropic Wilson~\cite{Beane:2009py} calculations, respectively.
\label{fig:a0_NN}
}
\end{figure}

Deuteron is a bound state of two nucleons in the $^3S_1$ and $I=0$ channel.
Nucleon-nucleon scattering in this channel and also in the $^1S_0$ channel
was first studied in quenched QCD~\cite{{Fukugita:1994na},{Fukugita:1994ve}}. 
This work was followed by a partially-quenched mixed action~\cite{Beane:2006mx}
and $N_f = 2+1$ anisotropic Wilson action~\cite{Beane:2009py} simulations.
Extraction of nuclear force between two nucleons has been investigated in
quenched and 2+1 flavor QCD~\cite{{Ishii:2006ec},{Aoki:2009ji},{Aoki:2008hh}}. 
Results for the scattering lengths $a_0$ from these studies are summarized in 
Fig.~\ref{fig:a0_NN}. The scattering lengths in the two 
channels are almost identical in each group. 
The results, however, have large discrepancies between the groups.
An even more problematic issue is that the absolute value 
of the lattice results 
is much smaller than the experimental values, $a_0 = 23.7$ fm 
and $a_0 = -5.47$ fm for the $^1S_0$ and $^3S_1$ channels, respectively.
The lattice results do not show 
strong dependence on the pion mass at the region where the calculations
were carried out, $m_\pi \simgt 0.3$ GeV.  In order to explain 
the experimental values, the scattering lengths have to vary significantly 
when calculations near the physical quark mass are carried out in future.
We should also note that all these studies 
assumed that the deuteron state is not bound for 
the heavy pion mass employed in the calculations.

Recently not only two-baryon systems but also
three-baryon systems have been investigated using lattice QCD.
NPLQCD Collaboration has tried a feasibility study 
of three-baryon systems focusing on the $\Xi^0\Xi^0 n$ and 
the $nnp$ (triton) channels.
They found both interactions to be 
repulsive~\cite{{Beane:2009gs},{Detmold:2009aa}}, 
which indicates that the triton is not bound for the parameters 
taken for the calculation.

In this conference several studies of two- and three-baryon systems 
were reported. 
HALQCD Collaboration studied (i) the energy dependence
of the nuclear force~\cite{Murano:2010aa}, (ii) the nuclear force
in the flavor SU(3) limit~\cite{Inoue:2010aa}, (iii) extraction 
of the two-baryon forces in a coupled channel with 
the variational method~\cite{Ishii:2010aa,Sasaki:2010aa}, 
and (iv) an exploratory study of extraction of the three-nucleon
force~\cite{Doi:2010aa}.
In multi-meson systems, NPLQCD Collaboration proposed
a recursion relation approach for multi-meson correlation 
functions~\cite{{Detmold:2010au},{Detmold:2010aa}} to
largely reduce the computational cost of the correlation functions.

\section{Computational issues with nuclei}
\label{Computational issues with nuclei}

There are several computational difficulties in the calculation 
of the multi-baryon bound state in lattice QCD. 
They are : 
1) exponential increase of statistical error,
2) factorial growth of fermion Wick contractions, and
3) identification of bound state.
While we avoid the first one by an unphysical heavy quark mass,
we propose solutions for the second and third problems.
Let us discuss each in turn. 

\subsection{Exponential increase of statistical error}

An estimate of the statistical noise to signal ratio 
for the correlation function of the nucleus consisting of 
$N_N$ nucleons is known~\cite{Lepage:1989hd} to be proportional to
\begin{equation}
\frac{1}{\sqrt{N_{\rm meas}}}\,
{\rm exp}\left(N_N\left[
m_N - \frac{3}{2}m_\pi\right]t
\right),
\end{equation}
where $m_\pi$ and $m_N$ are the masses of the pion and nucleon, respectively,
$N_{\rm meas}$ is the number of measurement, and $t$ is the separation
between the source and sink time slices.
The statistical error exponentially increases as
the number of nucleon increases as well as when the quark mass decreases.
We aim to treat helium nuclei in this work, so that $N_N$ is
fixed to four and three for $^4$He and $^3$He channels, respectively.
Since our main aim is to explore nucleus calculations, 
and since the difficulty of controlling statistical fluctuations toward 
the region of lighter pion mass is well known, 
we use the heavy quark mass corresponding to $m_\pi = 0.8$ GeV. 
Even then we had to carry out $O(10^3)$ measurements.

While this strategy would be acceptable for a feasibility test of
calculation of nucleus, we need novel methods to solve this
problem for a more realistic calculation
near the physical quark mass. We leave this task in future.

\subsection{Factorial growth of Wick contractions}

Another computational problem for multi-nucleon systems
is a factorially large number of Wick contractions of quark-antiquark fields 
required for evaluations of the nucleus correlation functions. 
A naive counting would give $(2N_p + N_n)!(2N_n + N_p)!$ for a nucleus composed of $N_p$ protons
and $N_n$ neutrons, which quickly becomes prohibitively large beyond 
three-nucleon systems, {\it e.g., } 2880 for $^3$He and 518400 for $^4$He.

This number, however, contains equivalent contractions under the permutation 
symmetry in terms of the protons or the neutrons in the interpolating operator.
We can reduce the computational cost by avoiding the redundancy.
In the case of the $^4$He nucleus which consists of the same number 
of protons and neutrons, the isospin symmetry 
also helps us reduce the necessary contractions.
After a scrutiny of the remaining equivalent contractions
by a computer we find that only 1107 (93) 
contractions are required for the $^4$He ($^3$He) nucleus correlation function. 
We have made a numerical test that the result with the reduced 
contractions reproduces the one with the full contractions on a
configuration.

For an additional technique to save the computational cost 
of the nucleus correlation functions,
we make a block of three quark propagators 
where a nucleon operator with zero spatial momentum is constructed
in the sink time slice.
In this procedure we can incorporate the permutation symmetry
of two up (down) quarks in a proton (neutron) sink operator.
This is a simple trick to calculate $2^{N_N}$ contractions simultaneously. 
We also prepare several combinations of the two blocks 
which are useful for the construction of the nucleus correlators.

\subsection{Identification of bound state}

A general issue with numerical calculations for exploring 
bound state formation is to distinguish the physical binding energy
from the energy shift due to the finite volume effect 
in the attractive scattering system~\cite{{Luscher:1986pf},{Beane:2003da},{Sasaki:2006jn}}.
The problem is made more difficult for nuclei because 
the binding energy $\Delta E$ of the nucleus consisting 
of $N_N$ nucleons with the mass $m_N$ is very tiny compared 
with the mass $M$ of the nucleus: 
$\Delta E / M \sim O(10^{-3})$ with $\Delta E = N_N m_N-M$.

One way to solve the problem is to investigate the volume dependence
of the measured energy shift:
In the attractive scattering system 
the energy shift is proportional to $1/L^3$
at the leading order in the $1/L$ expansion~\cite{Luscher:1986pf,Beane:2007qr}, 
while the physical binding energy remains at a finite value 
at the infinite spatial volume limit.

If the volume is not large enough, it is difficult to distinguish
a constant from a $1/L^3$ behavior in the energy shift.
Thus, in our simulation we employ large volumes as much as possible,
and choose three spatial extents corresponding to 3.1, 6.1 and 12.3 fm. 
The largest two volumes are much larger than those employed in current numerical simulations. 
They should provide sufficient room for the helium nuclei.

\section{A quenched calculation of Helium nuclei}
\label{A quenched calculation of Helium nuclei}

\subsection{Simulation details}

We carry out calculations on quenched configurations generated with the 
Iwasaki gauge action~\cite{Iwasaki:1983cj} 
% <================= Do NOT forget to insert correct reference!!!!
at $\beta = 2.416$ whose
lattice spacing is $a=0.128$ fm determined with $r_0=0.49$ fm 
as an input~\cite{AliKhan:2001tx}.
We employ the HMC algorithm with the 
Omelyan-Mryglod-Folk integrator~\cite{{Omelyan:2003om},{Takaishi:2005tz}}.
The step size is
chosen to yield reasonable acceptance rate presented 
in Table~\ref{tab:conf_meas}.
We take three lattice sizes, 
$L^3\times T = 24^3 \times 64$, $48^3 \times 48$ and $96^3 \times 48$, 
to investigate the spatial volume dependence of the energy 
difference between the
ground state of the nucleus channel and the free multi-nucleon states.
The physical spatial extents are 3.1, 6.1 and 12.3 fm, respectively.

\begin{table}[!t]
\centering{
\begin{tabular}{cccccc}\hline\hline
$L$ & $N_{\mathrm{conf}}$ & $N_{\mathrm{meas}}$ 
& accept.(\%)
& $m_\pi$ [GeV] & $m_N$ [GeV]\\
\hline
24 & 2500 &  2 & 93 & 0.8000(3) & 1.619(2) \\
48 &  400 & 12 & 93 & 0.7999(4) & 1.617(2) \\
96 &  200 & 12 & 68 & 0.8002(3) & 1.617(2) \\\hline\hline
\end{tabular}
}
\caption{
Number of configurations ($N_{\rm conf}$), 
number of measurements on each configuration ($N_{\rm meas}$),
acceptance rate in the HMC algorithm,
pion mass ($m_\pi$) and nucleon mass ($m_N$).
\label{tab:conf_meas}
}

\end{table}
We use the tadpole improved Wilson action 
with $c_{\mathrm{SW}} = 1.378$~\cite{AliKhan:2001tx}.
As discussed in the previous section, 
since it becomes harder to obtain a reasonable signal-to-noise ratio at
lighter quark masses for the multi-nucleon system, 
we employ a heavy quark mass at $\kappa = 0.13482$ which gives
$m_\pi = 0.8$ GeV for the pion mass and $m_N = 1.6$ GeV for the nucleon mass. 
Statistics are increased by repeating the measurement of 
the nucleus correlation functions
with the source points in different time slices on each configuration.
The numbers for the configurations and the measurements on each configuration
are summarized in Table~\ref{tab:conf_meas}.
We separate 100 trajectories between each measurement 
with $\tau=1$ for the trajectory length.
The errors are estimated by the jackknife analysis choosing 200 
trajectories for the bin size.

The quark propagators are solved with
the periodic boundary condition 
in all the spatial and temporal directions,
and using the exponentially smeared source
\begin{equation}
q^\prime(\vec{x},t) = \sum_{\vec{y}} A\, e^{-B|{\vec x} - \vec{y}|} q(\vec{y},t)
\end{equation}
after the Coulomb gauge fixing.
$q$ is the quark field at the source time slice and $A, B$ are the smearing
parameters.
On each volume we employ two sets of the  
smearing parameters: $(A,B) = (0.5,0.5)$ and $(0.5,0.1)$ 
for $L=24$ and $(0.5,0.5)$ and $(1.0,0.4)$ for $L=48$ and 96.
Effective mass plots with different sources, which are shown later, 
help us confirm the ground state in the nucleus channel. 
Hereafter the first and the second smearing parameter 
sets are referred to as "$S_{1,2}$", respectively.

The interpolating operator for the proton is defined as 
$p_\alpha = \varepsilon_{abc}([u_a]^tC\gamma_5 d_b)u_c^\alpha$
where $C = \gamma_4 \gamma_2$ and $\alpha$ and $a,b,c$ are the Dirac index and
the color indices, respectively.
The neutron operator $n_\alpha$ is obtained 
by replacing $u_c^\alpha$ by $d_c^\alpha$ 
in the proton operator.
To save the computational cost
we use the nonrelativistic quark operator, in which the Dirac index
is restricted to upper two components.

The $^4$He nucleus has zero total angular momentum and positive parity
$J^P = 0^+$ with the isospin singlet $I = 0$. 
We employ the simplest $^4$He interpolating operator with the 
zero orbital angular momentum $L=0$, and hence $J=S$ with $S$ being
the total spin.
Such an operator was already given long time ago in Ref.~\cite{Beam:1967zz},
\begin{equation}
^4\mathrm{He} = 
\left( \overline{\chi}\eta - 
\chi \overline{\eta} \right) / \sqrt{2},
\end{equation}
where 
\begin{eqnarray}
\chi &=&  ( [+-+-] + [-+-+] - [+--+] - [-++-] )/2,\\
\overline{\chi}  &=& ( 
[+-+-] + [-+-+] + [+--+] + [-++-] - 2 [++--] - 2 [--++] 
)/\sqrt{12}
\end{eqnarray}
with $+/-$ being up/down spin of each nucleon. 
$\eta, \overline{\eta}$ are obtained
by replacing $+/-$ in $\chi, \overline{\chi}$ by $p/n$ for the isospin.
Each nucleon in the sink operator is projected to have zero spatial
momentum.

We also calculate the correlation function of 
the $^3$He nucleus whose quantum numbers are 
$J^P=\frac{1}{2}^+$, $I = \frac{1}{2}$ and $I_z = \frac{1}{2}$.
We employ the interpolating operator in Ref.~\cite{Bolsterli:1964zz},
\begin{equation}
^3{\rm He} = 
\left(
\left|p_- n_+ p_+ \right\rangle
-
\left|p_+ n_+ p_- \right\rangle
+
\left|n_+ p_+ p_- \right\rangle
-
\left|n_+ p_- p_+ \right\rangle
+
\left|p_+ p_- n_+ \right\rangle
-
\left|p_- p_+ n_+ \right\rangle
\right)/\sqrt{6},
\end{equation}
with the zero momentum projection on each nucleon in the sink operator.

\begin{figure}[!t]
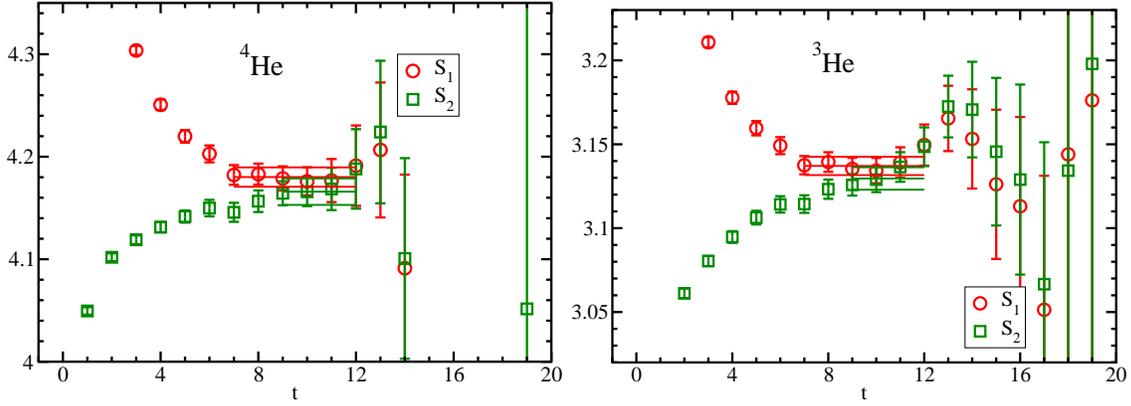

\includegraphics*[angle=0,width=0.49\textwidth]{Figs/eff_He.eps}
\includegraphics*[angle=0,width=0.49\textwidth]{Figs/eff_He3.eps}
\caption{
Effective masses for $^4$He (left) and $^3$He (right) correlation functions 
with $S_1$ (circle) and $S_2$ (square) 
sources at spatial extent of $6.1$ fm.
Fit results with one standard deviation error band are expressed 
by solid lines.
\label{fig:eff_He}
}
\end{figure}

\subsection{$^4$He channel}

Let us first present the results for the $^4$He channel.
The left panel of figure~\ref{fig:eff_He} shows the effective mass plots of 
the $^4$He nucleus correlators with the $S_{1,2}$ sources 
on the (6.1 fm)$^3$ spatial volume.
We find clear signals up to $t\approx 12$, beyond which
statistical fluctuation dominates.
The effective masses with the different sources show a reasonable
agreement in the plateau region. 
The consistency is also shown in the exponential fit results in
the plateau region as presented by the solid lines in the figure.

\begin{figure}[!t]
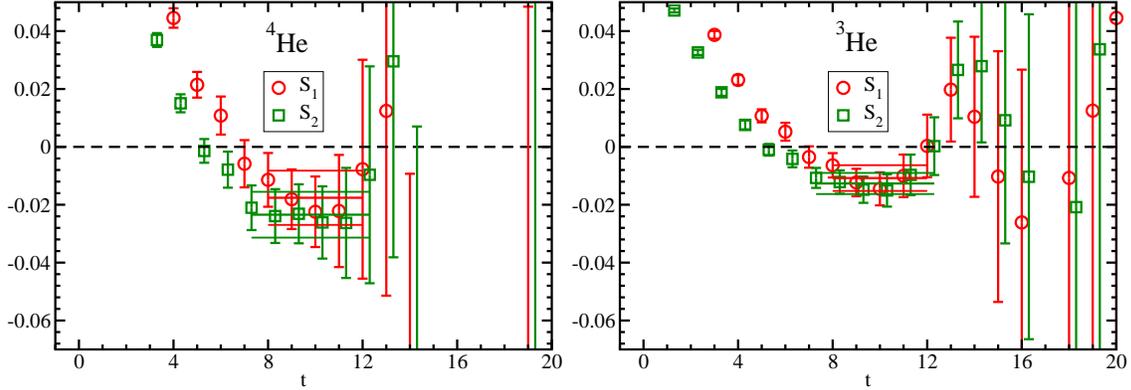

\includegraphics*[angle=0,width=0.49\textwidth]{Figs/eff_R_He.eps}
\includegraphics*[angle=0,width=0.49\textwidth]{Figs/eff_R_He3.eps}
\caption{
Effective energy shifts for $^4$He (left) and $^3$He (right) 
channels in a convention of $-\Delta E_L^{\mathrm{eff}}$
with $S_1$ (circle) and $S_2$ (square) sources 
at spatial extent of $6.1$ fm.
Square symbols are slightly shifted to positive direction in horizontal 
axis for clarity.
Fit results with one standard deviation error band are expressed 
by solid lines.
\label{fig:eff_R_He}
}
\end{figure}

In order to determine the energy shift $\Delta E_L$ precisely, 
we define the ratio of the $^4$He nucleus correlation function divided by
the fourth power of the nucleon correlation function,
\begin{equation}
R_{^4\mathrm{He}}(t) = \frac{G_{^4\mathrm{He}}(t)}{(G_N(t))^4},
\end{equation}
where $G_{^4\mathrm{He}}(t)$ 
and $G_N(t)$ are obtained with the same source.
The effective energy shift is extracted as
\begin{equation}
-\Delta E_L^{\mathrm{eff}} = \ln \left( \frac{R(t)}{R(t+1)} \right),
\label{eq:delta_E}
\end{equation}
once the ground states dominate in both of the correlators.
In the left panel of Fig.~\ref{fig:eff_R_He} we present time dependence of 
$-\Delta E_L^{\mathrm{eff}}$ for the $S_{1,2}$ sources, both of which
show negative values 
beyond the error bars in the plateau region of $8 \le t \le 11$.
Note that this plateau region is reasonably consistent 
with that for the effective mass
of the $^4$He nucleus correlators in the left panel of Fig.~\ref{fig:eff_He}.
The signals of $-\Delta E_L^{\mathrm{eff}}$ 
are lost beyond $t\approx 12$ because of 
the large fluctuations in the $^4$He nucleus correlator.
We determine $\Delta E_L$ by exponential fits of the ratios in 
the plateau region, $t=8-12$ for $S_1$ and 
$t=7-12$ for $S_2$, respectively.
We estimate a systematic error of $\Delta E_L$
from the difference of the central values of the fit results with
the minimum or maximum time slice changed by $\pm 1$.

\begin{figure}[!t]
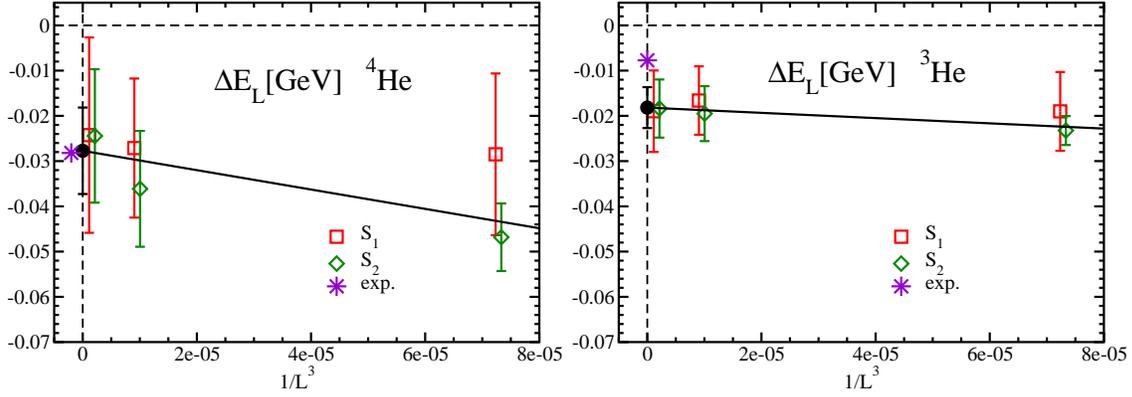

\includegraphics*[angle=0,width=0.49\textwidth]{Figs/dE_h.eps}
\includegraphics*[angle=0,width=0.49\textwidth]{Figs/dE_h3.eps}
\caption{
Spatial volume dependences for $-\Delta E_L = M - N_N m_N$ in GeV units
for $^4$He (left) and $^3$He (right) nuclei with $S_1$ (open square) 
and $S_2$ (open diamond) sources.
Statistical and systematic errors are added in quadrature.
Diamond symbols are slightly shifted to positive direction in horizontal axis
for clarity. Extrapolated results to the infinite spatial 
volume limit (filled circle) 
and experimental values (star) are also presented.
\label{fig:dE_He}
}
\end{figure}

Table~\ref{tab:dE} summarizes the numerical values
of the energy shift $\Delta E_L$ at three spatial volumes, 
where the statistical and systematic errors are presented in 
the first and second parentheses, respectively.
The volume dependence of $\Delta E_L$
is plotted as a function of $1/L^3$ in the left panel of 
Fig.~\ref{fig:dE_He}. 
The errors in the figure are evaluated from the statistical and 
systematic errors added 
in quadrature. In the following discussions in this subsection 
we use the combined error.
The results for the $S_{1,2}$ sources are consistent within the error bars.
We observe little volume dependence for $\Delta E_L$ indicating 
a bound state, rather than the $1/L^3$ dependence expected for a 
scattering state, for the ground state in the $^4$He channel.

The physical binding energy $\Delta E$ defined 
in the infinite spatial volume limit
is extracted by a simultaneous fit of the data for the $S_{1,2}$ sources 
employing a fit function of $\Delta E + C/L^3$ with $\Delta E$ and $C$ 
free parameters.
The $1/L^3$ term is added to allow for contamination of scattering states.
A systematic error is estimated from the difference of the central values
of the fit results using the data with the different fit ranges
in the determination of $\Delta E_L$.
The result for $\Delta E$ is 0.0180(62)
in lattice units, which 
is 2.9 $\sigma$ away from zero as shown in the left panel of 
Fig.~\ref{fig:dE_He}.
We also try a pure bound state fit allowing for an exponentially small 
finite size correction:
$\Delta E$ and $\Delta E + C_1 e^{-C_2 L}$ with 
$\Delta E$ and $C_{1,2}$ free parameters.
We find all the results are in agreement with reasonable values of $\chi^2$.

Based on these analyses we conclude that the ground state of the measured 
four-nucleon system is bounded.  
An encouraging finding is that $\Delta E$ = 27.7(9.6) MeV 
with $a^{-1}=1.54$ GeV 
agrees with the experimental value of 28.3 MeV. 
However, we do not intend to stress the consistency because
our calculation is performed at the unphysically heavy 
pion mass, $m_\pi = 0.8$ GeV, and the electromagnetic interactions and
the isospin symmetry breaking effects are neglected.

\begin{table}[!t]
\centering{
\begin{tabular}{ccccc}\hline\hline
$L$ & \multicolumn{4}{c}{$\Delta E_L$ [MeV]}\\
 & $^4$He($S_1$) & $^4$He($S_2$) & $^3$He($S_1$) & $^3$He($S_2$) \\
\hline
24 & 28(14)(11) & 46.8(7.3)(1.6) & 19.0(6.3)(6.0) & 23.2(3.2)(0.5)\\
48 & 27(14)(05) & 36(12)(04)     & 16.6(6.9)(3.2) & 19.5(5.6)(2.3)\\
96 & 24(18)(12) & 24(14)(03)     & 19.0(7.6)(4.9) & 18.4(6.1)(1.9)\\
$\infty$ & \multicolumn{2}{c}{27.7(7.8)(5.5)} &
\multicolumn{2}{c}{18.2(3.5)(2.9)}\\\hline\hline
\end{tabular}
}
\caption{
\label{tab:dE}
Energy shifts for $^4$He and $^3$He channels on each spatial volume. 
Extrapolated results to the infinite spatial volume limit 
are also presented. The first and second errors are
statistical and systematic, respectively.
}
\end{table}

\subsection{$^3$He channel}

The results of effective mass and the effective energy shift for
the $^3$He channel with the $S_{1,2}$ sources
are shown in the right panel of Figs.~\ref{fig:eff_He}
and ~\ref{fig:eff_R_He}, respectively.
The statistical error is slightly smaller than those for the $^4$He
channel.
We determine $\Delta E_L$ for the $^3$He channel as in the $^4$He channel,
whose results are presented in the right panel of 
Fig.~\ref{fig:dE_He} and Table~\ref{tab:dE}. 
The trend of the volume dependence is similar to the $^4$He channel case.
A simultaneous fit of the data for the $S_{1,2}$ sources 
with a fit function of $\Delta E + C/L^3$
yields a finite value of $\Delta E= 18.2(4.5)$ MeV, with the combined errors
as in the $^4$He channel, in the infinite volume 
limit.  This means the existence of a bound state
in the $^3$He channel. 
Our result for $\Delta E$, however, is about twice larger than 
the experimental value of 7.72 MeV. 
The main reason could be the heavy pion mass employed in this calculation.

As an alternative way to view this result, we compare the binding energies 
normalized by the atomic number: $\Delta E / N_N = 6.9(2.4)$ MeV and 
6.1(1.5) MeV for
the $^4$He and $^3$He nuclei, respectively.    
At our unphysically heavy pion mass, the three and four nucleon systems 
do not show the experimental feature that the binding is stronger for 
$^4$He than for $^3$He. 
We consider that this is mainly caused by the heavy quark mass
used in this calculation.

\section{Toward further progress}
\label{Toward further progress}

We have addressed the issue of nuclear binding for the case of $^4$He and $^3$He 
nuclei.  We have shown that the current computational techniques and 
resources allow us to tackle this issue.  Albeit in quenched QCD and for 
unphysically heavy pion mass, we are able to extract evidence for 
the bound state nature of the ground state and the binding energies for 
these nuclei. 

It is encouraging that our results for the binding energies 
and the experimental values are of same order of magnitude.
There are several issues which need clarification, however.
Our conclusion for $^3$He seems to disagree with the recent result
of NPLQCD Collaboration~\cite{{Beane:2009gs},{Detmold:2009aa}}.
While the two simulations have been carried out under different 
parameters, {\it e.g.,} quark mass, number of flavors, and volumes, 
further work is needed to obtain a consensus in this channel.
Furthermore we should gain a deeper understanding on how the helium nuclei 
forms a bound state at such a heavy quark mass.
Study of the nuclear force extracted from the wave function
and looking at the simplest nucleus, deuteron,
might help to understand our results.

A future direction 
of primary importance is to investigate the quark mass dependence
of the binding energies of the nuclei.
There are several model studies of the quark mass dependence of the
nuclear binding energies~\cite{Flambaum:2007mj} which suggest that
the quark masses play an essential role in a quantitative understanding
of the binding energies. 

Another important issue is development of a strategy to calculate
nuclei with larger atomic numbers.
The number of Wick contractions quickly diverges
as the atomic number increases,
even if the redundancies are removed with various symmetries and techniques
employed in this work.
At this conference, NPLQCD Collaboration presented a set of 
recursion relations~\cite{{Detmold:2010au},{Detmold:2010aa}} 
for correlation functions for the $n$-meson system, and also
for more complex systems with multi-species, such as $n$-$\pi$ and $m$-$K$ 
systems.
Similar recursion relations in multi-baryon systems might solve 
the problem.  We leave it to future work.

\section*{Acknowledgments}

We would like to thank the organizers of Lattice 2010
for the invitation of this presentation.
Numerical calculations for the present work have been carried out
on the HA8000 cluster system at Information Technology Center
of the University of Tokyo and on the PACS-CS computer 
under the ``Interdisciplinary Computational Science Program'' of 
Center for Computational Sciences, University of Tsukuba. 
We thank our colleagues in the PACS-CS Collaboration for helpful
discussions and providing us the code used in this work.
This work is supported in part by Grants-in-Aid for Scientific Research
from the Ministry of Education, Culture, Sports, Science and Technology 
(Nos. 18104005, 18540250, 20105002, 21105501, 22244018).

\bibliography{yamazaki}

\end{document}